\documentclass{aa}  

\usepackage{graphicx}
\usepackage{txfonts}
\usepackage{multirow}
\usepackage{subcaption}
\usepackage{threeparttable}
\usepackage{mathtools}
\usepackage{nicematrix}
\usepackage{booktabs, caption, makecell}
\begin{document}

   \title{ALMA polarimetry of radio-quiet AGNs}

   \author{E.~Shablovinskaia\inst{1,2}
          \and
          C.~Ricci\inst{1,3}
          \and 
          R.~Paladino\inst{4}
          \and
          A.~Laor\inst{5}
        \and
        C-S.~Chang\inst{6}
          \and
          D.~Belfiori\inst{4}
          \and 
          T.~Kawamuro\inst{7}
          \and
          E.~Lopez-Rodriguez\inst{8}
          \and 
          D.~J.~Rosario\inst{9}
          \and 
           S.~Aalto\inst{10}
          \and 
          M.~Koss\inst{11,12}
          \and 
        R.~Mushotzky\inst{13,14}
          \and 
          G.~C.~Privon\inst{15,16,17}
          }

   \institute{Instituto de Estudios Astrofísicos, Facultad de Ingeniería y Ciencias, Universidad Diego Portales, Av. Ejército Libertador 441, Santiago, Chile\\
\email{elena.shablovinskaia@mail.udp.cl}
\and Max-Planck-Institut für Radioastronomie, Auf dem Hügel 69, Bonn D-53121, Germany
\and Kavli Institute for Astronomy and Astrophysics, Peking University, Beijing 100871, China
\and INAF, Istituto di Radioastronomia di Bologna, via Piero Gobetti 101, 40129 Bologna, Italy
\and Physics Department, Technion, Haifa 32000, Israel
\and Joint ALMA Observatory, Avenida Alonso de Cordova 3107, Vitacura, Santiago 7630355, Chile
\and Department of Earth and Space Science, Osaka University, 1-1 Machikaneyama, Toyonaka 560-0043, Osaka, Japan
\and Kavli Institute for Particle Astrophysics \& Cosmology (KIPAC), Stanford University, Stanford, CA 94305, USA
\and School of Mathematics, Statistics and Physics, Newcastle University, Newcastle upon Tyne, NE1 7RU, UK
\and Department of Space, Earth and Environment, Chalmers University of Technology, SE-412 96 Gothenburg, Sweden
\and Eureka Scientific, 2452 Delmer Street Suite 100, Oakland, CA 94602-3017, USA
    \and Space Science Institute, 4750 Walnut Street, Suite 205, Boulder, Colorado 80301, USA
\and Department of Astronomy, University of Maryland, College Park, MD 20742, USA 
    \and Joint Space-Science Institute, University of Maryland, College Park, MD 20742, USA
    \and National Radio Astronomy Observatory, Charlottesville, VA 22903, USA
    \and Department of Astronomy, University of Florida, Gainesville, FL, 32611, USA
    \and Department of Astronomy, University of Virginia, Charlottesville, VA, 22904, USA}

   \date{Received September 15, 1996; accepted March 16, 1997}

% \abstract{}{}{}{}{} 
% 5 {} token are mandatory
 
  \abstract
   {The compact mm emission ubiquitously found in radio-quiet active galactic nuclei (RQ AGN) exhibits properties consistent with synchrotron radiation from a small region ($\leq$1 light day) and undergoing self-absorption below $\sim$100~GHz. Several scenarios have been proposed for its origin, including an X-ray corona, a scaled-down jet, or outflow-driven shocks, which can be tested via mm polarimetry. In the optically thin regime, synchrotron emission is expected to show polarization up to $\sim$70\%, but disordered magnetic fields and Faraday rotation reduce this to a few percent for jets and outflows, while an X-ray corona is likely to result in complete depolarization. To investigate this, we conducted the first ALMA Band 3 full-polarization observations of three RQ AGN -- NGC~3783, MCG~5--23--16, and NGC~4945. No polarized signal was detected in any of the AGN, with an upper limit of 0.5--1.5\%, supporting the X-ray corona scenario. However, we detected a compact source with 17\% polarization in NGC~3783, 20 pc away from the AGN, co-spatial with the mm and narrow-line outflow, likely linked to a shock propagating through the outflowing material. Additionally, combining our data with archival ALMA observations, we found typical mm variability in RQ AGN by a factor of 2; however, the sparsity of the data prevented a more detailed analysis of the total flux variability.}

   \keywords{techniques: polarimetric -- galaxies: active --  submillimeter: galaxies
               }

   \maketitle
%
%-------------------------------------------------------------------

\section{Introduction}

The class of radio-quiet active galactic nuclei (AGN) represents $\sim$90\% of all observed AGN {\citep[e.g.][]{Kellermann89,Ivezic02}. However, drawing a clear boundary between radio-quiet (RQ) and radio-loud (RL) AGN }can be challenging in practice, as even RQ objects often exhibit faint radio emission. Its origin is currently under active debate \citep[e.g.][]{panessa19}, with various mechanisms proposed, including nuclear star formation, thermal free-free emission, and compact synchrotron sources. Notably, radio emission in AGN has been found to correlate with X-ray emission \citep{laor08,panessa15}, similar to what is observed in coronally active stars \citep{guedel93}.

The nature of high-frequency radio emission, particularly in the millimeter (mm) range, remains even more enigmatic. Historically, the spectral region near 100--300~GHz has received limited attention \citep[e.g.][]{hickox18}, but recent studies over the past decade have revealed unresolved mm emission located at the center in the majority of RQ AGN ($\sim$92\% according to \citealt{Kawamuro22}, or 94$^{+3}_{-6}$\% according to \citealt{ricci23}). This mm emission demonstrates an even tighter correlation with X-ray emission than lower-frequency radio does \citep{Kawamuro22,ricci23}. \citet{Inoue18} found that RQ AGN exhibit an excess of mm emission that cannot be extrapolated from lower-frequency radio data. The spectral shape of this excess aligns well with synchrotron radiation impacted by self-absorption, with a turnover frequency near 100~GHz. This turnover frequency suggests a very compact emitting region; modeling provides estimates of the source size of $\sim$40–50 gravitational radii ($R_{\rm g}$), a scale similar to that of the accretion disk corona \citep{Inoue18}. The compactness of the mm-emitting region is further supported by observations of rapid, significant variability \citep{shab24}, which imply an upper limit of about 1 light-day in size and effectively rule out emission mechanisms like thermal emission from dust or free--free emission, which cannot account for such rapid variations.

Given the combined characteristics -- compact size, mm/X-ray correlation, and spectral shape -- the most likely emission mechanism appears to be synchrotron radiation produced by the population of non-thermal electrons associated with the X-ray corona in AGN. However, the lack of correlated variability between the mm and X-ray bands \citep{behar20,petrucci23}, even in high-resolution, high-cadence observations \citep{shab24}, raises further questions about the nature of the mm emission. While the synchrotron mechanism remains the most plausible origin, the exact source is still under debate. In addition to compact size, the source should contain a magnetic field and a population of non-thermal electrons. Beyond the X-ray corona, where these conditions are expected to occur \citep[see][]{Inoue14}, other possible origins include scaled-down jets or shocks within AGN outflows \citep[see][for further discussion]{Kawamuro22,shab24}.

To date, various observational approaches, including spectral shape analysis, variability studies, and X-ray/mm correlations, have been employed to reveal the properties of the mm-emitting source and definitively pinpoint its origin \citep[e.g.,][]{behar15,Inoue14,Kawamuro23}.
Given that synchrotron emission should be initially highly polarized, investigating the polarimetric properties of mm emission shows a promising avenue for uncovering its nature. In this manuscript, we explore the polarimetric characteristics of compact mm emission in RQ AGN, beginning with an examination of the expected polarization signatures from different synchrotron sources. To test these predictions, we conducted the first-ever mm polarimetric survey of a sample of RQ AGN, all of which exhibit compact mm emission \citep{Kawamuro23,ricci23}, observed at frequencies close to the expected synchrotron turnover. This study aims to shed new light on the mechanisms driving mm emission in RQ AGN and contribute to a more comprehensive understanding of its origins.

\section{Expected levels of mm polarization} \label{sec_depol}

The most likely mechanism responsible for mm emission in RQ AGNs is synchrotron radiation from a compact source. While the most plausible scenario is the X-ray corona, other possibilities include a scaled-down jet or a shock in the outflow. Observing mm emission near the turnover frequency or at higher frequencies may allow us to detect synchrotron radiation in the optically thin regime. 
In cases where synchrotron emission occurs in an optically thin medium with a strong, ordered magnetic field, the polarization degree (PD) depends solely on the power-law energy distribution of non-thermal electrons, which can be described by $dn_{\rm e}^{\rm nt} / dE \propto E^{-\gamma}$, where $E$ is the electron energy and $\gamma$ is the electron index. Then the linear polarization should be observed with a degree equal to:
\begin{equation*}
    {\rm PD}_{\rm int} = \frac{3\gamma + 3}{3\gamma+7},
\end{equation*}
so, for typical values of $\gamma$ between 2 and 3, the PD is up to $\sim$70\% \citep{leroux,RL79}. In this scenario, the polarization degree is wavelength-independent and oriented perpendicular to the magnetic field\footnote{{In the optically thick regime, the polarization angle is rotated by 90$^\circ$ relative to the optically thin case, and polarization remains wavelength-independent, with the intrinsic PD given by ${\rm PD}_{\rm int} = \frac{3}{6\gamma+13}$, yielding values around 10--12\% for $\gamma = 2-3$ \citep{pach77}} .}.

\subsection{Beam depolarization}

In reality, synchrotron emission experiences significant depolarization. The primary cause is beam depolarization, which occurs due to variations in the magnetic field direction within a source that is more compact than the angular resolution limit. This effect can arise from specific magnetic field configurations, such as a toroidal field, or extreme disorder in the field structure. In the latter case, the synchrotron source can be modeled as consisting of $N_{\rm c}$ cells with randomly oriented magnetic fields, leading to a decrease in the total PD as $\propto 1/\sqrt{N_{\rm c}}$. Since all emission sources considered here are expected to be non-uniform due to turbulence \citep[as in jets, e.g.,][]{marscher14}, magnetic reconnection (in jets, e.g. \citealt{lyutikov03}, or coronae, e.g. \citealt{MF01}), or other instabilities, depolarization should inevitably occur. However, its exact impact cannot be reliably estimated without knowing the structure of the emitting region.

\subsection{Faraday depolarization}

Another important mechanism of depolarization, especially in the radio band, is Faraday depolarization. The polarization plane of the emission passing through a medium of thermal electrons in the presence of a magnetic field is rotated. The same rotation occurs in the case of the external Faraday effect, where the emission source and the magnetized medium are spatially separated, and for the internal Faraday effect, where they are co-spatial. Faraday rotation can be described by the rotation measure (RM):
\begin{equation*}
 {\rm RM} = 2.62 \times 10^{-19}  \int_{0}^{R} n_{\rm e}^{\rm th} B dr \, \, \,  \, \, \, {\rm (rad \,m^{-2})}   
\end{equation*}
which depends on the thermal electron density $n_{\rm e}^{\rm th}$ (in cm$^{-3}$) and the magnetic field strength $B$ (in Gauss), with only the magnetic field component along the line of sight being relevant. In general, $n_{\rm e}^{\rm th}$ and $B$ are functions of the path $r$ (in cm) traveled by the radiation within a medium of size $R$. The resulting polarization angle rotation is then given by
\begin{equation*}
    \Delta \chi = \lambda^2 {\rm RM},
\end{equation*}
where $\lambda$ is the observed wavelength.

In the case of internal Faraday rotation, synchrotron radiation is produced in different regions of a finite-sized source, where the medium contains a magnetic field. As a result, the radiation traverses different path lengths within the source, leading to varied rotation angles. This causes depolarization of the observed total emission \citep{gardner66}. The depolarization factor, i.e., the ratio of observed to intrinsic polarization ${\rm PD}_{\rm obs}/{\rm PD}_{\rm int}$, is a function of RM and $\lambda^2$, which depends on the source geometry and medium uniformity. The simplest model is a uniform, symmetric slab \citep{burn66,sokoloff98}, which can be applied both to the corona \citep[e.g.][]{raginski16} and jets and outflows \citep[e.g.][]{Yushkov24}. In this case, depolarization is given by:
\begin{equation} \label{depol}
    {\rm PD}_{\rm obs} / {\rm PD}_{\rm int} = \left| \frac{\sin(2 \lambda^2{\rm RM})}{2\lambda^2{\rm RM}} \times \exp^{2i(\lambda^2{\rm RM})}\right|.
\end{equation}
Complete depolarization occurs when $\lambda^2$RM $= k\pi$ for integer $k$, although this condition varies for different types of slabs  \citep[see][]{sokoloff98}. In some slab types and in a spherically symmetric source, full depolarization does not occur at all (see \citealt{burn66}). However, for most configurations (except for anomalous "reversed" depolarization due to specific magnetic field configurations, \citealt{sokoloff98}), the observed polarization will decrease significantly at sufficiently large $\lambda^2$RM.

\subsection{Faraday rotation in mm emitting sources}

Since our observation wavelength is fixed (100~GHz, or 3~mm), we can roughly estimate RM {for internal Faraday rotation} for the three mm-emitting synchrotron sources we consider.

\textit{Jet.} In general, a parsec-scale jet is not suitable for describing the mm excess in RQ AGN due to its size, as a larger synchrotron-emitting region corresponds to a lower turnover frequency, which for jets is assumed to be much lower than 100~GHz. However, a compact jet may still be present in RQ AGN, as we observe in radio maps at lower frequencies, and its material could depolarize mm emission. Faraday rotation and depolarization have indeed been observed in jets, including with ALMA, with characteristic RMs in the range of $10^3-10^5$~rad m$^{-2}$ \citep[e.g.,][]{goddi21, peng24}. Even with relatively low electron densities ($\sim$10$^{-2}$~cm$^{-3}$) and magnetic field strengths of $0.1-10$~G \citep{OSullivan09, hovatta19}, these RMs result in significant polarization suppression. For example, at 3 mm, RM is typically up to $4 \times 10^4$~rad m$^{-2}$, and in extreme cases, RMs can reach $5 \times 10^5$~rad m$^{-2}$ at 1.3 mm \citep{hovatta19}, reducing synchrotron polarization to $\sim$1-2\%.

\textit{Outflow.} When considering AGN warm ionized outflows, we assume they consist mainly of thermal electrons, which produce unpolarized emission due to free-free radiation processes. However, shocks can occur in these outflows, compressing both the material and the magnetic field, thus increasing their density. Such processes may induce a low level of polarization, around 7\% \citep[as in NGC~1068,][]{lr20}. However, due to the compact size of the shock fronts, they are unlikely to produce significant internal Faraday rotation. The thermal medium in the outflow could contribute to depolarization, but this effect is expected to be minor: with an average particle density of $\sim$10$^{2}-10^3$~cm$^{-3}$ \citep{davies20}\footnote{In certain AGN outflows, densities can reach 10$^{4}$--10$^{5}$~cm$^{-3}$ or more. Such high-density outflows are usually associated with ultrafast outflows \citep[e.g.][]{xu24} and ultraluminous infrared galaxies or compact radio sources \citep[e.g.][]{holt11,santoro18}, none of which apply to the sources in our sample.} and a relatively weak magnetic field\footnote{The typical strength of the large-scale magnetic field in galaxies is up to a few dozen microgauss. Here, we assume an upper limit of $B$ of a few milligauss, as observed in magnetized clouds of the Milky Way \citep[see][for a review]{MF_review}.}, even a parsec-scale outflow would have an RM around $\sim$10$^5$~rad m$^{-2}$. Consequently, we may still observe a polarization level of a few percent in these cases.

%Footnote 2, the density in AGN outflows can actually reach 10^10, which is what you derive in BALQSO. In fact, it likely drops as 1/r^2, which is what radiation pressure compression produces, and as indicated by the observed outflow ionization parameter. 

\textit{Corona.} The corona in AGNs is highly compact, with a size estimated to be less than 0.001~pc ($\sim$3$\times 10^{15}$~cm) from microlensing measurements of mm-emitting source \citep{rybak_eas}, or confined to the region 1--100~$R_{\rm g}$ ($\sim$3$\times 10^{12}$ -- 3$\times$10$^{14}$~cm) {for slab geometry in modeling of the X-ray data via ray tracing \citep{Marinucci19,Gianolli23}.} Corona consists mostly of hot ($T_{\rm e} \sim 10^9$~K) thermal electrons with only few per cent \citep[$\sim$1--5\%,][]{Inoue08,Inoue14} of non-thermal electron population. The electron density $n_{\rm e}^{\rm th}$ within the corona is thought to be less than $10^{10}$~cm$^{-3}$ \citep{haardt93}. It can be estimated from the optical depth $\tau$ and corona size $L$: $n_{\rm e}^{\rm th} \approx \tau / (\sigma_{\rm T}R)$, where $\sigma_{\rm T} = 6.65 \times 10^{-25}$~cm$^{2}$ is the Thomson cross-section. Concerning the median estimate of $\tau \approx 0.25$ \citep{ricci18}, we can estimate the density as $\sim$10$^{9}$--10$^{11}$~cm$^{-3}$. For $B \approx 1 - 10$~G \citep{laor08}, RM reaches up to $\sim$10$^{11}$~rad m$^{-2}$, leading to extreme depolarization according to Eq.~(\ref{depol}), with an undetectable polarization degree less than 10$^{-8}$\%.

Additionally, for such extreme RM values, depolarization also occurs due to the finite bandwidth of the observations. If the rotation is so large that within the bandwidth the polarization angle $\chi$ rotates by more than $\pi$, the measured polarization will effectively average to zero. Thus, the maximum measurable RM in observations at frequency $\nu = c / \lambda$ with a bandwidth $\Delta \nu$ is constrained by:
\begin{equation*}
    {\rm RM} \leq \frac{\pi}{2 \lambda^2 \Delta \nu / \nu}. 
\end{equation*}
For observations at 100~GHz with $\Delta \nu =$ 2~GHz, the maximum measurable RM $\approx 9 \times 10^6$~rad m$^{-2}$.

Thus, if there is a jet or outflow with a compact source of synchrotron emission passing through its medium, we are likely to detect polarization at a fairly low level, on the order of a few percent. In the case of a corona, however, the observed polarization will be strictly zero.

%M87 jet: \citep{peng24}, $4\times 10^4$ rad m-2, 0.9\%
%extreme RM: \citep{hovatta19}, $5\times 10^5$ rad m-2, 1.8\%

\begin{table*}[!ht] 
  \centering
    \caption{Sample of RQ AGN. AGN types, redshifts ($z$), distance ($D$), SMBH masses ($M_{\rm SMBH}$), and column densities ($N_{\rm H}$) are taken from \citet{koss_catalogue} and \citet{ricci17a}; references for the observed inclination angles $i$ are given in the caption below: X-ray luminosities ($L_{\rm 14-150~keV}$) are taken from \citet{ricci23}; Eddington ratios ($\lambda_{\rm Edd}$) are calculated using X-ray luminosity and considering bolometric correction of 8.48 (see text below).}
\begin{threeparttable}[]
\begin{tabular}{lcccccccc}
\hline
Name        & Type & $z$     & $D$ & $\log M_{\rm SMBH}$    & $\log N_{\rm H}$    & $i$ & $\log L_{\rm 14-150~keV}$  & $\log \lambda_{\rm Edd}$  \\
            &    &   & (Mpc)  & ($M_\odot$)  &  (cm$^{-2}$) & (deg) & (erg s$^{-1}$) & \\
\hline
\noalign{\smallskip}
NGC 3783    & 1  & 0.009 & 38.5 & 7.37  & 20.49 & 15.0\textsuperscript{\textdagger} & 43.45 & $-1.09$ \\
\noalign{\smallskip}
MCG 5--23--16 & 1.9  & 0.008 & 36.2 & 7.65  & 22.18 & 41$^{+9}_{-10}$\textsuperscript{\textdaggerdbl} & 43.44  & $-0.48$ \\
\noalign{\smallskip}
NCG 4945    & 2  & 0.002 & 3.7 & 6.15 & 24.60 & 75$\pm$2\textsuperscript{\textsection} & 41.63  & $-1.69$ \\
\noalign{\smallskip}
\hline
\end{tabular}
     \begin{tablenotes}
     \item[] \textsuperscript{\textdagger} \citet{Fischer13}.
     \item[] \textsuperscript{\textdaggerdbl} \citet{Serafinelli23}.
     \item[] \textsuperscript{\textsection} \citet{henkel18}
   \end{tablenotes}
    \end{threeparttable}%
\label{tab:sample}
\end{table*}

\section{Sample, observations and data reduction}

\begin{figure}
    \centering
    \includegraphics[width=0.90\linewidth]{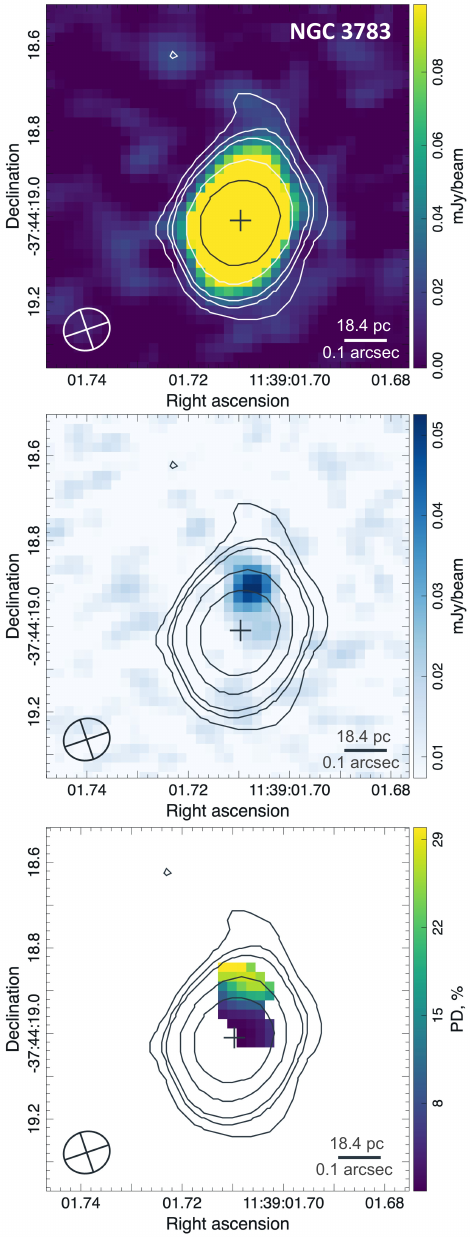}
    \caption{ALMA observations of NGC~3783. \textit{Upper panel}: total intensity with the contours corresponding to the 2, 4, 6, 18, 50$\sigma$ levels (1$\sigma$ = 0.007~mJy/beam) in the box $\sim$0.9$\times$0.9$''$. The location of the AGN according to \textit{Gaia} DR3 \citep{gaia_dr3} is marked with a black cross. \textit{Middle panel}: polarized intensity with the overplotted total intensity contours. \textit{Bottom panel}: polarization degree in per cent. The areas with a signal-to-noise ratio $<$ 5 on the total intensity map are masked.}
    \label{fig:3783}
\end{figure}

To investigate the polarimetric properties of mm emission, we selected RQ AGN from a volume-limited sample (<50~Mpc) which has been observed before at 100~GHz by ALMA with <100\,mas resolution \citep{ricci23}. All of these AGN showed unresolved nuclear emission with resolutions down to 0\farcs05--0\farcs1. From the sample from \citet{ricci23}, we chose the three brightest sources at 100~GHz, selecting objects of different AGN types and covering the widest possible range of column densities. This selection was designed to ensure a variety of inclination angles, allowing us to investigate both potential absorption effects in external AGN structures and to observe different magnetic field projections. Finally, we chose the following three RQ AGN: NGC~3783, MCG~5--23--16, and NGC~4945. The properties of the AGN are summarized in Table~\ref{tab:sample}.

The ALMA Band 3 observations were conducted in October 2023 (2023.1.01517.S; PI C. Ricci), during the ALMA long baseline configuration (C-8), with the longest baseline extending to 8.5~km. The observation dates, time spent on the sources, and the beam sizes of the obtained images are summarized in Table~\ref{tab:obs}. The spectral setup was the default frequency setup for polarization continuum observations, maximizing the sensitivity, with four spectral
windows (1.985~GHz wide) divided into 64 channels, at 90.52, 92.48, 102.52, and 104.48~GHz.

Data processing was carried out using \texttt{CASA} version 6.5.4.9 and ALMA Pipeline version 2023.1.0.124 \citep{alma_pipeline}. The clean images were generated using CASA task \texttt{tclean} with weighting = briggs (robust=0.5). According to the ALMA Proposer’s Guide, the systematic flux error for Band 3 observations is 5\%. Flux measurements from the polarization calibrators, as well as the check sources, yielded consistent results, justifying the application of a 5\% error margin in our analysis. Furthermore, the estimated degree and angle of polarization for the calibrators were consistent with the measurements from the AMAPOLA calibrator monitoring\footnote{\url{https://www.alma.cl/~skameno/AMAPOLA/}}.

For each epoch, we created maps of the Stokes parameters $I$, $Q$, $U$, and $V$.  Circular polarization was not detected in the maps and was therefore assumed to be zero, as it is typically weak at millimeter wavelengths. The polarized intensity was thus calculated as
\begin{equation*}
    P = \sqrt{Q^2 + U^2},
\end{equation*}
where $Q$ and $U$ are the Stokes parameters in units of flux. The polarization degree (PD) was then calculated as PD = $P/I \times 100\%$, where $I$ is the total intensity, and debiased following the method outlined in \citet{montier1,montier2}. In the cases when the polarization is undetected, we estimated the upper limit of the polarization degree: 
\begin{equation*}
    {\rm PD}_{\rm lim} = \frac{3 \times \sqrt{\sigma_Q^2 + \sigma_U^2 }}{I} \times 100\%, 
\end{equation*}
where $\sigma_Q$ and $\sigma_U$ are the standard deviation of the Stokes $Q$ and $U$, respectively, and $I$ corresponds to the peak total intensity, ensuring that ${\rm PD}_{\rm lim}$ is evaluated at the position of maximum signal-to-noise ratio in the image.

%--------------------------------------------------------------------
\section{Results}

\begin{figure}
    \centering
    \includegraphics[width=0.90\linewidth]{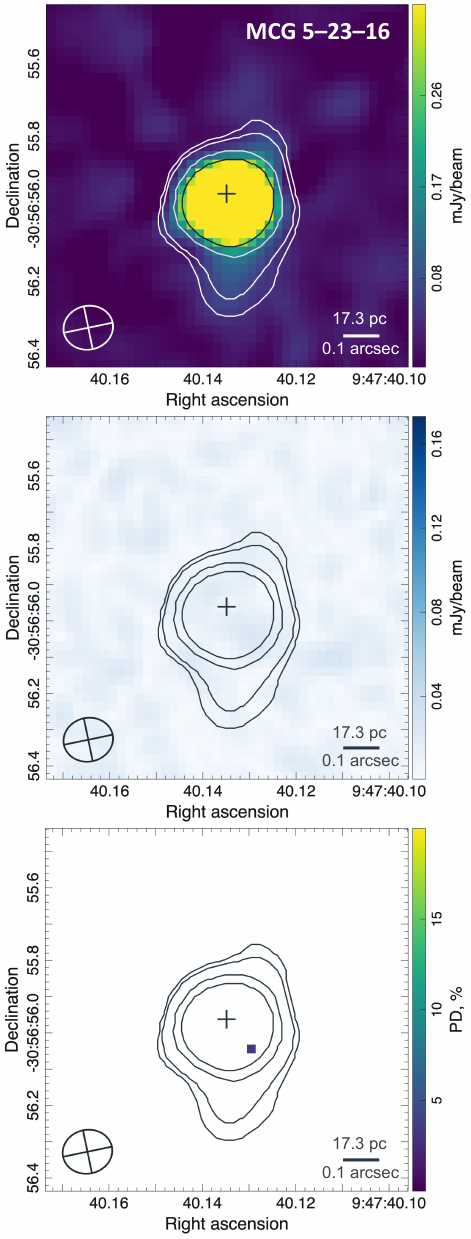}
    \caption{ALMA observations of MCG~5--23--16. \textit{Upper panel}: total intensity with the contours corresponding to the 4, 6, 19, 37$\sigma$ levels (1$\sigma$ = 0.010~mJy/beam) in the box $\sim$1$\times$1$''$. The location of the AGN according to \textit{Gaia} DR3 \citep{gaia_dr3} is marked with a black cross. \textit{Middle panel}: polarized intensity with the overplotted total intensity contours. \textit{Bottom panel}: polarization degree in per cent. The areas with a signal-to-noise ratio $<$ 5 on the total intensity map are masked.}
    \label{fig:mcg}
\end{figure}

\begin{figure}
    \centering
    \includegraphics[width=0.90\linewidth]{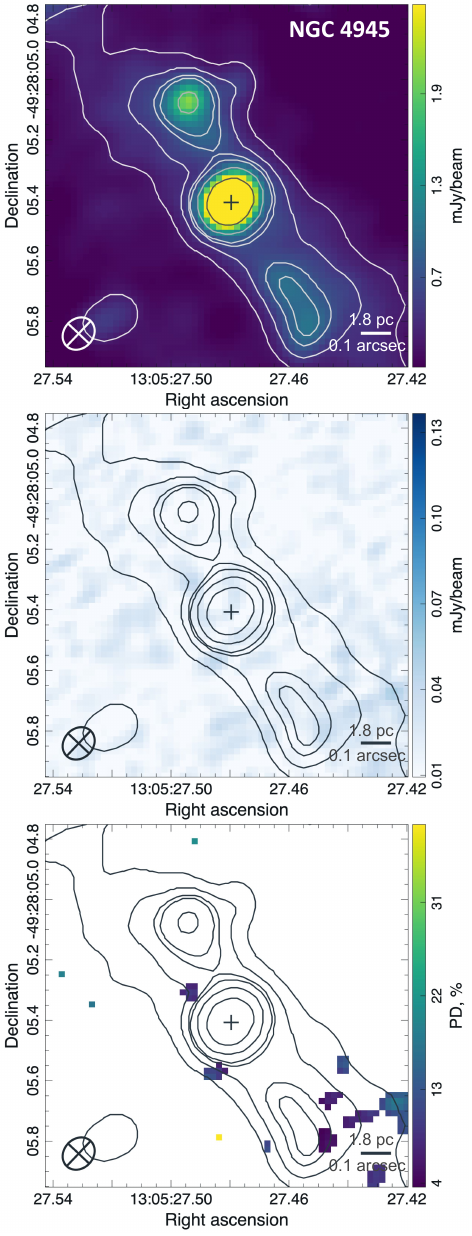}
    \caption{ALMA observations of NGC~4945. \textit{Upper panel}: total intensity with the contours corresponding to the 31, 46, 55, 92, 184$\sigma$ levels (1$\sigma$ = 0.016~mJy/beam) in the box $\sim$1.2$\times$1.2$''$. The location of the AGN, where mm emission peaks, is marked with a black cross. \textit{Middle panel}: polarized intensity with the overplotted total intensity contours. \textit{Bottom panel}: polarization degree in per cent. The areas with a signal-to-noise ratio $<$ 5 on the total intensity map are masked.}
    \label{fig:4945}
\end{figure}

\begin{table*}
\centering
\caption{ALMA Band 3 observation log. $^*$Due to insufficient parallactic angle coverage (51.12$^\circ$ instead of the desired 60.00$^\circ$) for the polarization calibrator on 2023$-$10$-$11, the observations were repeated on 2023$-$10$-$25.}
\begin{tabular}{lcccccc}
\hline
Name                           & Date~~          & $t$ & Beam size    & {rms $I$} & {rms $Q$} & {rms $U$ }               \\
                       &       & (min) &    & (mJy/beam) & (mJy/beam) & (mJy/beam)              \\
\hline 
\noalign{\smallskip}
NGC~3783                  & 2023--10--13~~   &     110           &   0\farcs109 $\times$ 0\farcs098   & 0.007 & 0.007 & 0.007                        \\
\noalign{\smallskip}
MCG~5--23--16             & 2023--10--08~~   &    59            &    0\farcs137 $\times$ 0\farcs121   & {0.010} & {0.010} & {0.010 }                      \\
\noalign{\smallskip}
\multirow{2}{*}{NGC~4945} & 2023--10--11$^*$ & 38             & 0\farcs123 $\times$ 0\farcs105 & {0.016} & {0.013} & {0.013} \\
                               & 2023--10--25~~ & 75             & 0\farcs140 $\times$ 0\farcs102 & {0.014} & {0.010} &
                               {0.010} \\
\noalign{\smallskip}
\hline
\end{tabular}
\label{tab:obs}
\end{table*}

\subsection{NGC~3783}

In the Seyfert 1 galaxy NGC~3783, we measured a peak mm flux of 2.01 $\pm$ 0.10~mJy from the central region, where the AGN is expected to reside, with no detectable variability over nearly two hours of observations. As the data were obtained with a significant on-source integration time and an angular resolution of ~0\farcs1 ($\sim$18.4~pc), we found that the mm continuum source appears slightly extended to the north (Fig.\ref{fig:3783}, upper panel), but this extended emission is faint, at the $\sim$0.1--0.2~mJy/beam level, and does not contribute substantially to the total flux, which remains dominated by the compact central source. The in-band spectral index of the AGN emission, calculated across four spectral windows assuming $F_\nu \propto \nu^{\alpha}$, was found to be $\alpha = -0.92 \pm 0.40$.

The maps of polarized intensity and PD (Fig.~\ref{fig:3783}, middle and bottom panel, respectively) showed no polarized signal in the AGN location, where also the mm intensity peaks, giving just an upper limit of PD$_{\rm lim} = 1.2$\%. 

Nevertheless, we detected polarized emission offset to the north from the AGN, along the extended structure. The maximum of the total intensity and the maximum of the polarized intensity are separated by $\sim$0\farcs11, which corresponds to the resolution limit of the data and a projected distance of $\sim$20.2~pc. This confirms that the polarization does not originate from the AGN itself. The offset mm emission exhibits a polarization degree of 17\% $\pm$ 2\%, with the polarization angle oriented at 10$^\circ$ $\pm$ 3$^\circ$. The in-band spectral index of total intensity in the region exhibiting polarized emission is extremely steep, $-3.6 \pm 2.4$, though the large uncertainty significantly limits the robustness of this measurement.

\subsection{MCG~5--23--16}

The observations of type 1.9 AGN MCG~5--23--16 revealed a slightly extended structure, with most of the intensity concentrated in the central compact source (Fig.~\ref{fig:mcg}, upper panel). We measured the flux 2.54 $\pm$ 0.13~mJy peaking in the position of the AGN, with no variability during the observing period. The mm emission showed an in-band spectral index $\alpha = -0.42 \pm 0.40$. The total intensity map indicates a slightly resolved structure, oriented along a position angle (PA) of about $-20^\circ$ to $189^\circ$ (north-south direction), with the southern lobe extending further, reaching $\sim$0\farcs3 ($\sim$52~pc). 
%and exhibits a significantly steeper spectral index of $\alpha \approx -2.3$.

The polarized intensity and PD maps are given in the middle and bottom panels of Fig.~\ref{fig:mcg}. No polarization was detected in this object, with an upper limit of PD$_{\rm lim} = 1.5$\%.

\subsection{NGC~4945}

NGC~4945, a Seyfert 2 galaxy with a nuclear starburst, is known for hosting numerous mm sources in its central region, previously identified as star clusters \citep{emig20}, inside the extended (>10$''$) mm continuum structure reproduced by combining dust thermal emission and free-free emission \citep{bendo16}. This complex structure was also observed in our Band\,3 observations. Additionally, we found a highly polarized mm source at a projected distance of $\sim$3\farcs4 from the nucleus, as reported in Shablovinskaya et al. (in prep.).
Although the surrounding material contributes significantly to the mm emission of the galaxy, the central compact AGN-associated source stands out as the brightest in the high-resolution data.

This galaxy was observed twice, with the sessions separated by 14 days. Despite formally insufficient parallactic angle coverage during the first session (51.12$^\circ$ instead of the desired 60.00$^\circ$) to reliably constrain instrumental polarization leakage and accurately recover the intrinsic polarization angle (see details in, e.g., \citealt{3c286,MV16}), it was deemed close enough to allow flux and polarization estimates. The measured fluxes were consistent across both epochs, with 9.14 $\pm$ 0.46~mJy in the first session and 8.82 $\pm$ 0.44~mJy in the second, aligning well with previous measurements \citep[8.1~mJy,][]{ricci23}. The spectral index remained stable as well, with $\alpha = -1.06 \pm 0.40$ and $\alpha = -1.29 \pm 0.40$ for the two epochs, respectively. 
Note that, despite exhibiting the highest mm flux among the objects in our sample, NGC~4945 has a mm luminosity that is $\sim$25 times lower, setting it apart from the others. This significant difference in luminosity likely causes the host galaxy of NGC~4945 to stand out prominently in our observations, while the other two AGNs appear almost point-like.

Fig.~\ref{fig:4945} presents the total intensity, polarized intensity, and PD maps from the first epoch. In the central region, coinciding with the AGN and the maximum mm flux, no polarized emission was detected. The second epoch yielded similar results, and thus, no additional images are shown. The estimated upper limits on polarization are low, with PD$_{\rm lim} = 0.6$\% and PD$_{\rm lim} = 0.5$\% for the two epochs, respectively.

\,

To summarize the measured properties of all AGN in the sample, we present fluxes and luminosities at 100~GHz, in-band spectral indices, and PD limits in Table~\ref{tab:sum}. Additionally, we provide 2--10~keV X-ray luminosities derived from \textit{Swift} observations taken within a few weeks of the ALMA data: 27 days later for NGC~3783 (obsid 00037255022), 19 days for MCG~5--23--16 (obsid 00030839036), and 38 days for NGC~4945 (obsid 00037266008).

\begin{table*}[!ht]
\centering
\caption{Summary of the measured properties of mm emission in the sample of RQ AGN.}
\begin{threeparttable}[]
\begin{tabular}{lccccc}
\hline
Name                      & Flux 100~GHz & log $L_{\rm 100~GHz}$ & $\alpha$        & PD$_{\rm lim}$ & log $L_{\rm 2-10~keV}$ \\
                          & (mJy)       & (erg s$^{-1}$)  &              & (\%)  & (erg s$^{-1}$)  \\
                          \hline
\noalign{\smallskip}
NCG 3783                  & 2.01 $\pm$   0.10    & 38.54 & $-0.92 \pm 0.40$   & 1.2  & 42.97 \\
\noalign{\smallskip}
MCG 5--23--16               & 2.54   $\pm$  0.13  & 38.59  &  $-0.42 \pm 0.40$             & 1.5  & 43.18 \\
\noalign{\smallskip}
\multirow{2}{*}{NGC 4945} & 9.14   $\pm$   0.46 & 37.13  & $-1.06 \pm 0.40 $ & 0.6 & \multirow{2}{*}{41.62\textdagger} \\
                          & 8.82    $\pm$  0.44  & 37.15  & $-1.29 \pm 0.40 $ & 0.5 \\
                          \hline
\noalign{\smallskip}
\end{tabular}
     \begin{tablenotes}
     \item[] \textsuperscript{\textdagger} For the Compton-thick AGN NGC~4945, the intrinsic X-ray luminosity was calculated assuming that only the flux normalization varied, using the normalization from \citet{ricci17a}.
   \end{tablenotes}
    \end{threeparttable}%
\label{tab:sum}
\end{table*}

%--------------------------------------------------------------------
\section{Discussion}

\subsection{Extended mm structure and polarization} \label{sec_ext}

The RQ AGN in our sample were selected to contain a compact, unresolved mm core. Our ALMA observations in full polarization mode with significant on-source time confirm that the central compact source still dominates in each case. However, due to the high sensitivity of these observations and a signal-to-noise ratio > 200, new faint (at the level of a few percent of the peak flux) details emerge in the total intensity maps, revealing that none of the sources appear exclusively compact anymore.

\citet{Kawamuro23} observed the same objects in ALMA Band 6, but noted extended structure only in NGC~4945, which, as discussed in Sec. 5.3 and referring to \citet{bendo16} and \citet{emig20}, is due to mm-bright star-forming regions and an extended thermal dust and free--free continuum component (3$''$ $\times$ 14$''$, \citealt{Kawamuro23}).
%referring to \citet{bendo16} and \citet{emig20}, the central region of this galaxy hosts both bright, compact mm sources associated with star-forming regions and an extended thermal dust and free-free continuum component (3$''$ $\times$ 14$''$, \citealt{Kawamuro23}). 
The other two sources in our sample, NGC~3783 and MCG~5--23--16, which previously appeared compact \citep{Kawamuro23,ricci23}, now show faint extended components at the 2--3$\sigma$ level, extending several dozen parsecs from the core (see Fig. \ref{fig:3783} and \ref{fig:mcg}). Although the spatially compact central emission and the extended mm emission overlap, without kinematic information on this faint extended component, we cannot confirm whether these structures originate from AGN activity.

NGC~3783 presents a particularly intriguing case. The extended mm emission in this galaxy was not reported by \citet{Kawamuro23} in their ALMA Band 6 observations. Furthermore, the ALMA archive lacks comparable observations with sufficient sensitivity and resolution to detect extended structures at the 0.1--0.2~mJy/beam level. The most important result is that our data reveal significant polarization in this region. Notably, the peak of the polarized emission is significantly offset -- by more than one beam size -- from the AGN position, suggesting that the polarized mm source is not directly associated with the close vicinity of the AGN. However, it appears that this polarization is co-spatial with the extended mm structure. Observations in other bands confirm the presence of an outflow in the same direction. MUSE observations with $\sim$0\farcs03 sky resolution indicate that the extended narrow-line region in NGC~3783 is elongated in this direction \citep{denBrok20}, a finding corroborated by \citet{gravity_3783}. This suggests that the extended mm emission may be part of an AGN-driven outflow, possibly radiating via free-free processes. In this scenario, the observed polarization could be associated with the outflow, where shocks produce synchrotron emission.

However, this hypothesis requires further investigation. If the polarization arises from a shock in the outflow, the observed PD $\sim$ 17\% is unexpectedly high. As discussed in Sec. 2, Faraday rotation and the turbulent nature of the shock are likely to depolarize the synchrotron radiation, typically resulting in PD values of only a few percent. Moreover, our estimate of the PD may represent a lower limit to the intrinsic polarization. The polarized source, with a total flux of $\sim$0.2~mJy, is located near the brighter AGN ($\sim$2~mJy), meaning that the observed total intensity could be contaminated by the neighboring AGN emission. This would lead to an overestimation of the total flux and, consequently, an underestimation of the PD, implying that the true intrinsic PD could be even higher, which would be even more difficult to reconcile with standard shock scenarios. Confirming this interpretation will require detailed modeling of the visibility data. Nonetheless, the detection of a $\sim$17\% polarized mm source in NGC~3783 already demonstrates not only the feasibility of detecting mm polarization in RQ AGNs, but also suggests that there is no significant Faraday depolarization screen on scales smaller than $\sim$20~pc.

\begin{figure}
    \centering
    \includegraphics[width=0.95\linewidth]{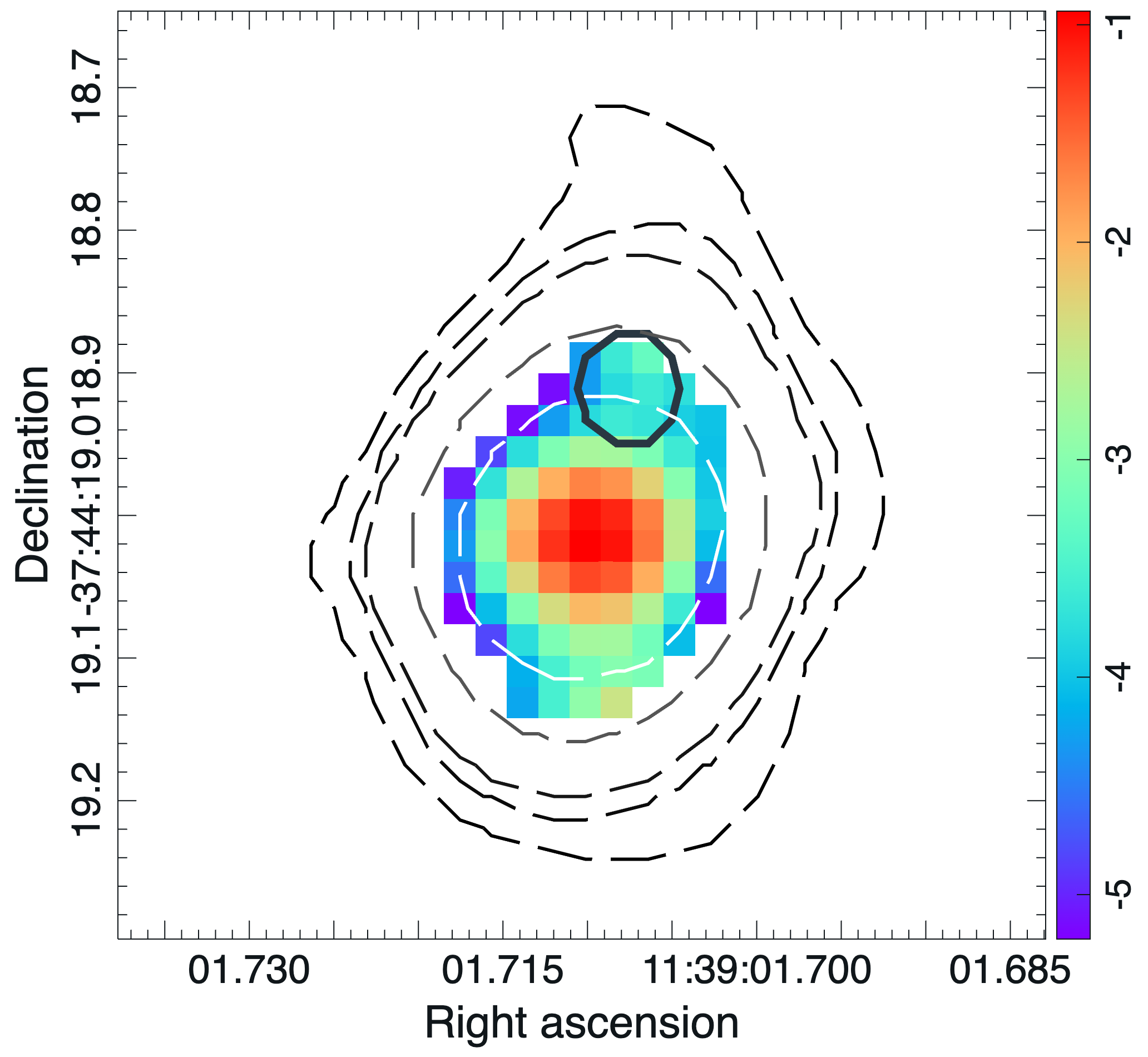}
    \includegraphics[width=0.95\linewidth]{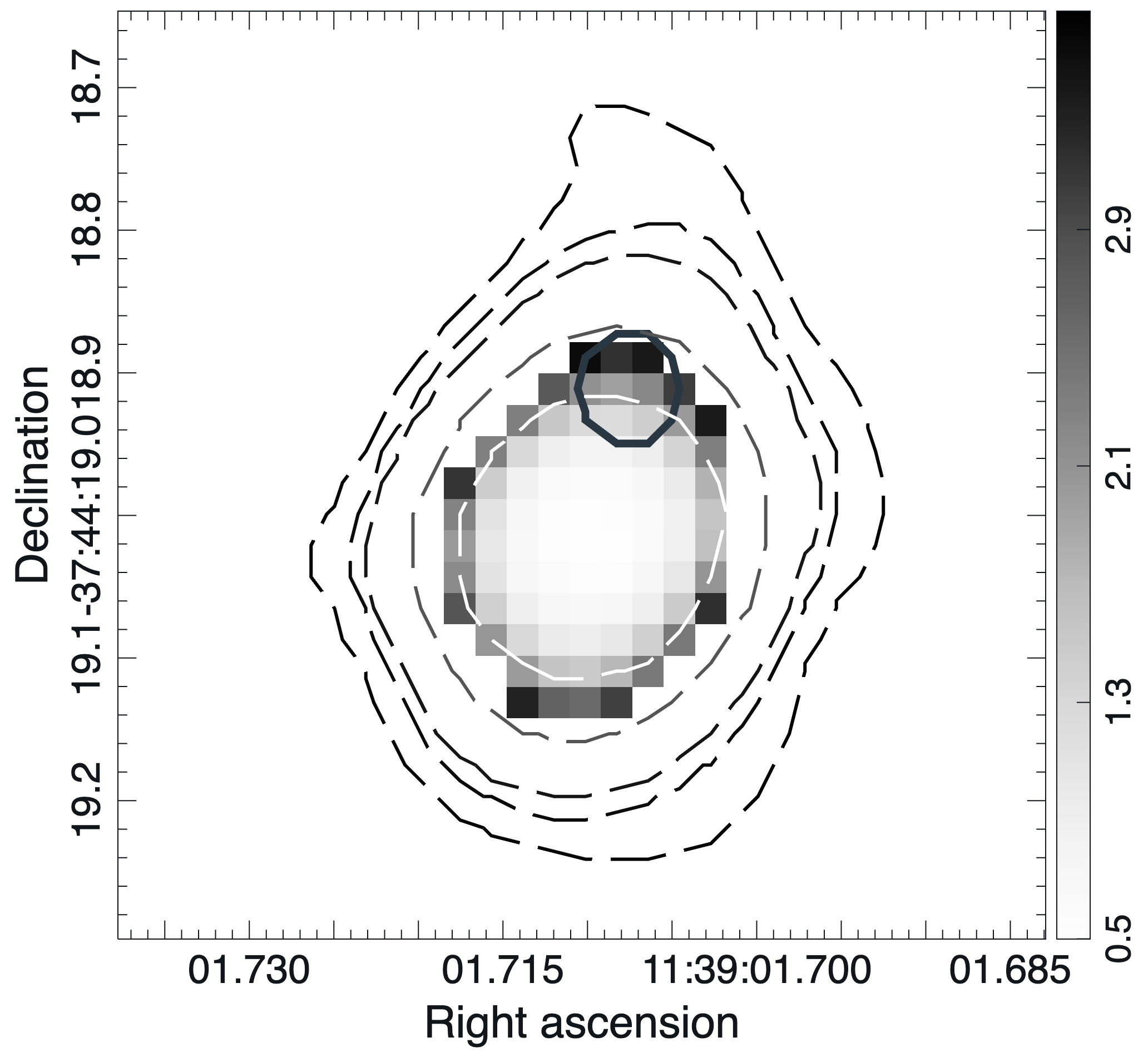}
    \caption{Map of the in-band spectral index $\alpha$ (upper panel) and its associated error (bottom panel). Dashed contours represent the total intensity (see Fig.~\ref{fig:3783}). The position of the polarized source is highlighted by a thick black contour. Note that the very steep slope in the outer regions is likely an artifact caused by the slightly narrower beam size at higher frequencies.}
    \label{fig:3783_spix}
\end{figure}

A similar challenge arises in estimating the spectral index of the polarized region. The maps of the in-band spectral index and its error are shown in Fig.~\ref{fig:3783_spix}. The maps are created for only pixels with a value exceeding 2$\sigma$ at all four frequencies. As seen in the figure, the spectral index at the AGN position is approximately $-$0.9. %In contrast, the polarized source located north of the AGN corresponds to a compact feature in the spectral index map, with an extremely steep in-band spectral index of $-$3.6 $\pm$ 2.4. This value is atypical for most known astrophysical emission mechanisms; however, it could be caused by photon absorption processes or by rapid electron cooling, where electrons lose energy with a restricted maximum energy limit. To better understand the nature of this emission, additional observations with broader spectral coverage in polarized light are essential.
In contrast, the polarized source located north of the AGN appears as a compact feature in the spectral index map, exhibiting an extremely steep in-band spectral index of $-3.6 \pm 2.4$. Although this value is unusually steep for most known astrophysical emission mechanisms, the large associated uncertainty limits a definitive interpretation. Nevertheless, such a steep slope, if real, could indicate rapid synchrotron cooling, where electrons lose energy before reaching high energies. To clarify the nature of this emission, further observations with broader spectral coverage in polarized light are crucial.

\begin{figure*}[h]

\begin{subfigure}{0.33\textwidth}
\includegraphics[width=0.9\linewidth]{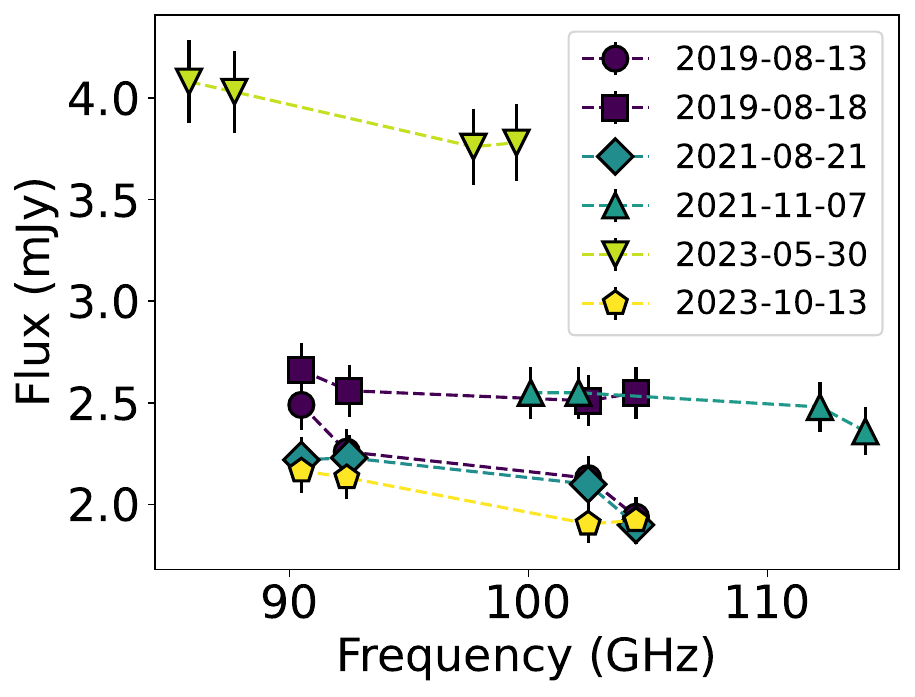} 
\caption{NGC~3783}
\label{fig:subim1}
\end{subfigure}
\begin{subfigure}{0.33\textwidth}
\includegraphics[width=0.9\linewidth]{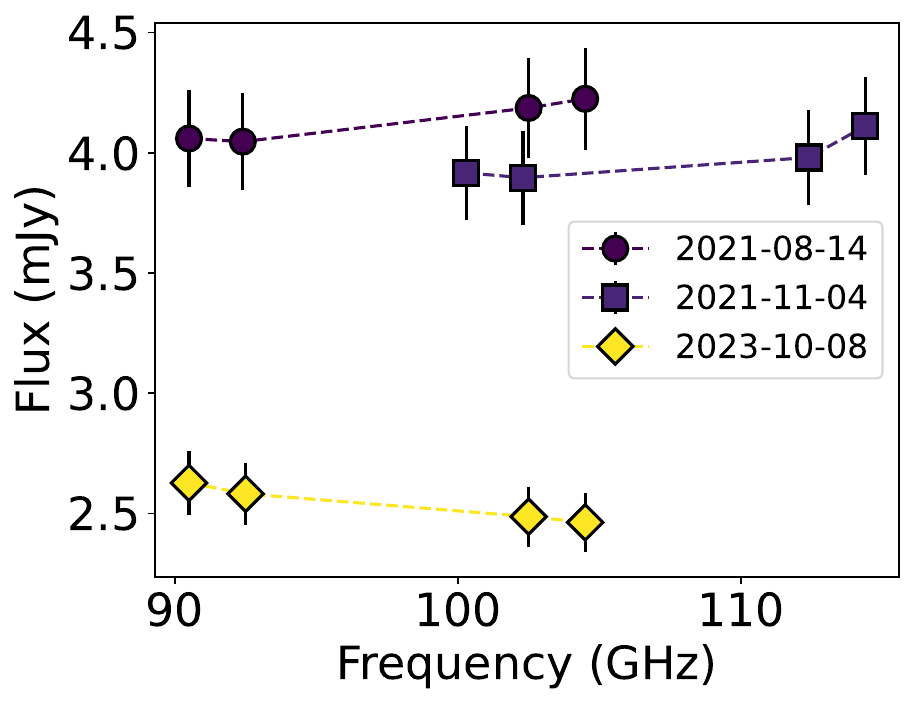}
\caption{MCG~5--23--16}
\label{fig:subim2}
\end{subfigure}
\begin{subfigure}{0.33\textwidth}
\includegraphics[width=0.9\linewidth]{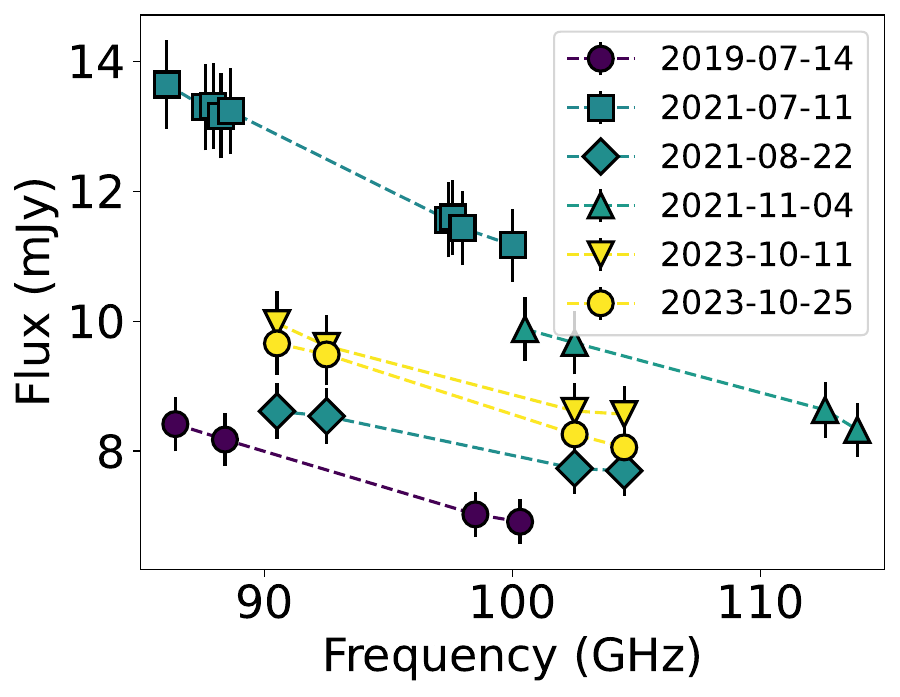}
\caption{NGC~4945}
\label{fig:subim2}
\end{subfigure}

\caption{Archival ALMA Band 3 data. Only observations with a resolution higher than 0\farcs3 were used. For all data, the flux error is assumed to be 5\%.}
\label{fig:arc_alma}
\end{figure*}

% \begin{figure*}[h]

% \begin{subfigure}{0.33\textwidth}
% \includegraphics[width=0.9\linewidth]{3783_SED_full.pdf} 
% \caption{NGC~3783}
% \label{fig:subim1}
% \end{subfigure}
% \begin{subfigure}{0.33\textwidth}
% \includegraphics[width=0.9\linewidth]{MCG_SED_full.pdf}
% \caption{MCG~5--23--16}
% \label{fig:subim2}
% \end{subfigure}
% \begin{subfigure}{0.33\textwidth}
% \includegraphics[width=0.9\linewidth]{4945_SED_full.pdf}
% \caption{NGC~4945}
% \label{fig:subim2}
% \end{subfigure}

% \caption{Archival radio and ALMA data. Only observations with a resolution higher than 0.3$''$ \textbf{[check]} were used. For all ALMA data, the flux error is assumed to be 5\%. The radio fluxes were taken from literature including errors.}
% \label{fig:arc_rad}
% \end{figure*}

\subsection{What causes depolarization of AGN mm emission?} \label{sec_what}

As it was discussed in Sec.~\ref{sec_depol}, the compact mm emission in the center of RQ AGN is highly likely to have a synchrotron origin. The spectral indices observed for the sources in our sample (see Table 3) also confirm this: NGC~3783 and NGC~4945 show $\alpha \approx -1$, consistent with optically thin synchrotron emission. MCG~5--23--16 has a less steep index, $\alpha \approx -0.4$, which in principle does not exclude a free-free origin. However, despite the strong evidence for synchrotron emission, we do not detect any significant polarization, suggesting that extreme depolarization must be at play. The key question is which mechanism -- non-uniform structure of the magnetic fields or Faraday rotation -- dominates.

As discussed in Sec. 2, all plausible sources of synchrotron mm emission in AGNs -- jets, shocks in outflows, and coronae -- are expected to be non-uniform. These regions are likely clumpy, and if we assume that the emitting area consists of $N_{\rm c}$ independent cells, the expected synchrotron polarization, $\sim$70\% can be suppressed as $\propto 1/\sqrt{N_{\rm c}}$. To reach even the highest measured upper limit of PD (1.5\%), the emission region would require $N_{\rm c} > 2000$ uncorrelated cells. Such a large number of randomly oriented emission zones would also smooth out any variability. However, as it was shown in recent studies, the mm emission in all observed RQ AGN are highly variable on the time-scales of days and shorter \citep{behar20,petrucci23,shab24,Michiyama24}. This makes this mechanism alone an unlikely explanation for the observed depolarization.

To produce depolarization, the magnetic field does not need to be completely disordered. Since the compact mm emission from AGN cores is unresolved, beam depolarization can play a significant role, e.g., if the synchrotron emission originates from electrons spiraling along magnetic loops. In this case, the polarization vectors rotate along the curved field lines, and when integrated over the unresolved structure, the net polarization is naturally diminished. The only way to overcome this effect is through extremely high-resolution observations. However, for coronal-scale structures, this would require angular resolutions at the microarcsec level, beyond the capabilities of even space-Earth interferometer RadioAstron \citep{radioastron} or the upcoming ngVLA.

Faraday depolarization is also an efficient depolarization mechanism. As previously discussed, even moderate Faraday rotation depths can significantly suppress polarization at mm wavelengths (see Sec.~\ref{sec_depol}). The efficiency of this effect depends on the electron density and magnetic field structure along the line of sight, which vary between different potential mm-emitting components. In jets and shocks in the outflows, the Faraday depth is expected to be lower than in the dense, magnetized corona. Given the observed polarization limits, a key argument in favor of the corona scenario is that it naturally provides the conditions for strong Faraday depolarization at mm wavelengths. 

In principle, this effect is testable: multi-frequency polarization measurements should reveal the characteristic $\lambda^{-2}$ dependence of PD, allowing estimates of RM. However, a non-detection in a narrow frequency window around 100~GHz is insufficient to constrain the origin of the depolarization. While the corona currently appears to be the most plausible source, confirming this requires polarization measurements at higher frequencies. For coronal emission expected to dominate in the 100--300~GHz range \citep[e.g.,][]{behar15}, a continued lack of polarization detection would imply RM values exceeding $\sim$10$^{8}$~rad~m$^{-2}$, significantly higher than those typically associated with jets or outflows -- and thereby strongly favoring the corona as the origin of the mm emission.

\subsection{Variability in mm} \label{sec_var}

In addition to polarization, variability analysis could provide crucial insight into the origin of mm emission. By examining variability timescales, correlations with other wavelengths, and the shapes of potential flares, one can compare these characteristics with predictions from physical processes that modulate synchrotron emission. While mm variability in RQ AGNs is not the primary focus of this paper, studying flux variations could help distinguish between different scenarios of its origin and, thus, the cause of depolarization. For this reason, we analyzed the flux of each source for possible variability.

First, we checked for intra-observational variability by re-imaging the observations and dividing them into time intervals of tens of minutes. No source exhibited significant flux changes beyond the measurement uncertainty (assumed to be 5\%). NGC~4945 also showed no detectable flux variations between the two epochs separated by 14 days. 

To examine variability on longer timescales, we compiled ALMA Band 3 data for all three sources. We selected observations with comparable angular resolution (less than 0\farcs3) to minimize contamination from nearby structures. The data are presented in Fig.~\ref{fig:arc_alma} and stored in Tables~\ref{arc_alma_3783}, \ref{arc_alma_MCG}, and \ref{arc_alma_4945} for NGC~3783, MCG~5--23--16, and NGC~4945, respectively. The observed flux variations between epochs were a factor of 1.9 in 136 days for NGC~3783, 1.6 in 703 days for MCG~5--23--16, and 1.7 in 42 days for NGC~4945. NGC~3783 and NGC~4945 showed changes in the in-band spectral index, which remained negative, varying between $-$0.26 $\pm$ 0.16 and $-$1.34 $\pm$ 0.39 in NGC~3783, and between $-$0.84 $\pm$ 0.06 and $-$1.35 $\pm$ 0.03 in NGC~4945. Surprisingly, archival data for MCG~5--23--16 revealed a positive spectral slope of $\sim$0.3. No correlation was found between flux and spectral index.

The predominantly negative spectral index in most observations supports the synchrotron origin of mm emission \citep{RL79}. The intensity of synchrotron emission depends on the energy density of the magnetic field ($\propto B^2$) and the kinetic energy of relativistic electrons, determined by their Lorentz factors and number density \citep[e.g.][]{essential_ra}. Unstable processes such as shock waves, plasma instabilities, and magnetic reconnection can accelerate electrons \citep[e.g.][]{part_acc}, thereby influencing synchrotron emission. The same processes also affect the local magnetic field energy, either increasing its density (e.g., in shock waves) or rapidly converting it into plasma heating and particle acceleration (e.g., in magnetic reconnection events).

If the observed variability in outflows is associated with shock waves, variability in the corona or relativistic jets is more likely driven by magnetic reconnection. In both cases, these processes heat the plasma, which could be expected to result in simultaneous increases in X-ray luminosity. Unfortunately, no simultaneous X-ray observations were available, preventing a direct comparison between X-ray and mm variability. However, in the mm regime, the shape and duration of a flare should differ depending on the underlying mechanism \citep[see][]{shab24}.

For a shock wave propagating through an outflow, one would expect a rapid rise followed by an exponential decline in intensity, similar to supernova light curves (see mm light curves simulations in \citealt{SN_mm_models}). The flare duration would depend on the shock propagation speed. Assuming the shock propagates through the AGN outflow, where the electron plasma temperature is $10^4-10^5$~K, the minimum possible speed is the local sound speed, $\sim$0.001$-0.003c$. Comparing this to our mm light curves, the most rapid observed flux drop was detected in NGC~4945, where the intensity decreased by nearly a factor of two over 40 days (between epochs 2021--07--11 and 2021--08--22). However, at such a low speed, the shock wave would travel less than a light-day in this time, which is significantly smaller than the expected outflow size.

In contrast, magnetic reconnection would produce a different flare profile. By analogy with solar coronal reconnection events, most of the released energy would be emitted in X-rays from the heated plasma, leading to a rapid rise followed by a prolonged decay in the X-ray band. The mm emission would increase due to the acceleration of non-thermal electrons, resulting in a sharp flare with a quick decline. The reconnection timescale depends on the Alfvén velocity in the medium. For an X-ray corona with a magnetic field of $1-10$~G and an electron number density of $\sim$10$^{10}$~cm$^{-3}$, the Alfvén speed is $\sim$0.01$-0.1c$. This suggests that a magnetic reconnection-driven flare should last no longer than $10-100$ days. However, our data for MCG~5--23--16 contradict this expectation: the source remained in a high state for at least four months in 2021.

Given the sparsity of the data, it is impossible to establish a definitive variability pattern. Flux variations by a typical factor of two are observed across all sources, but the large time gaps between observations hinder our ability to link high and low states. More regular monitoring is needed to draw meaningful conclusions about the variability mechanisms at play.

\begin{figure*}[h]

\begin{subfigure}{0.33\textwidth}
\includegraphics[width=0.95\linewidth]{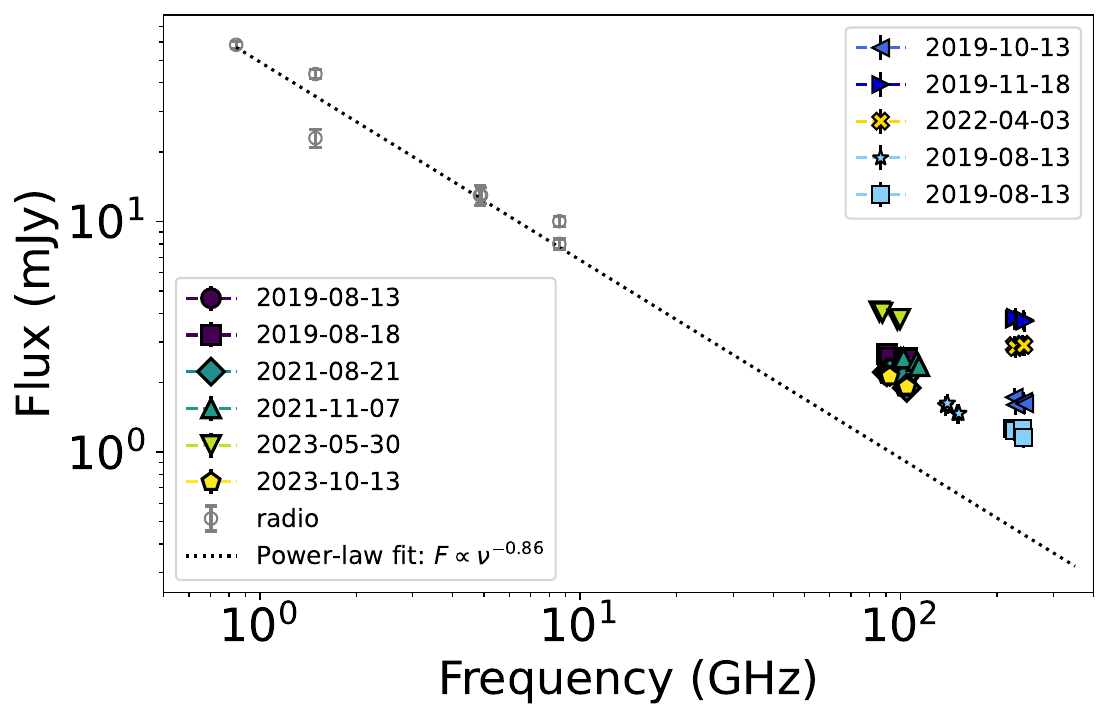} 
\caption{NGC~3783}
\label{fig:subim1}
\end{subfigure}
\begin{subfigure}{0.33\textwidth}
\includegraphics[width=0.95\linewidth]{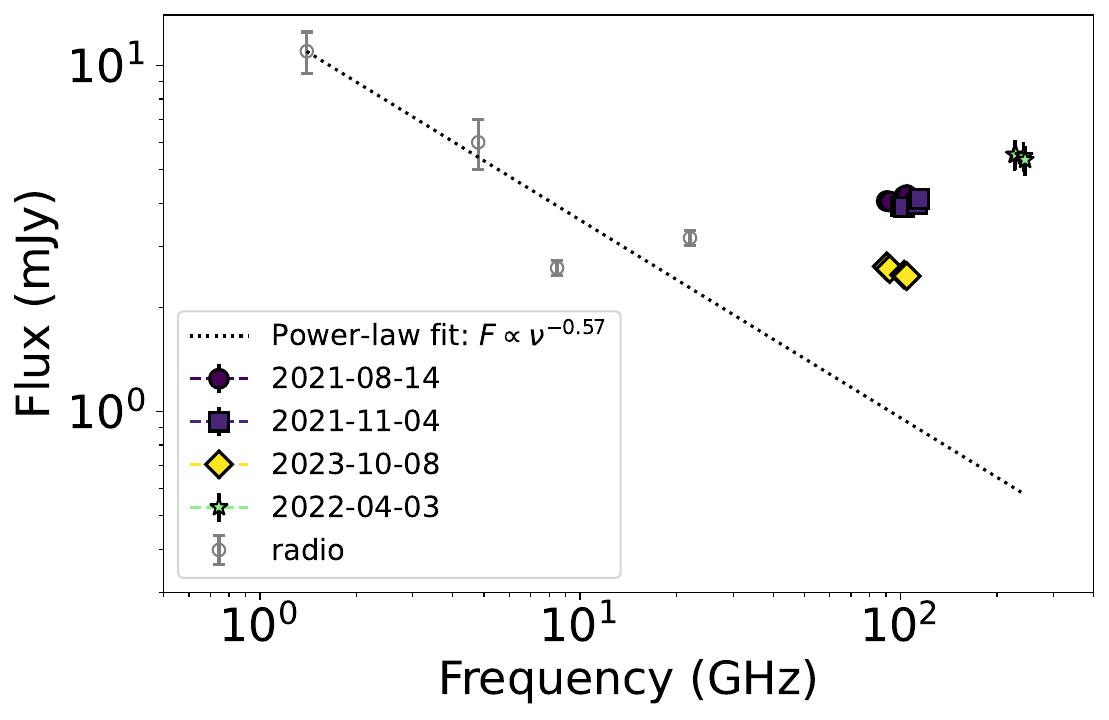}
\caption{MCG~5--23--16}
\label{fig:subim2}
\end{subfigure}
\begin{subfigure}{0.33\textwidth}
\includegraphics[width=0.95\linewidth]{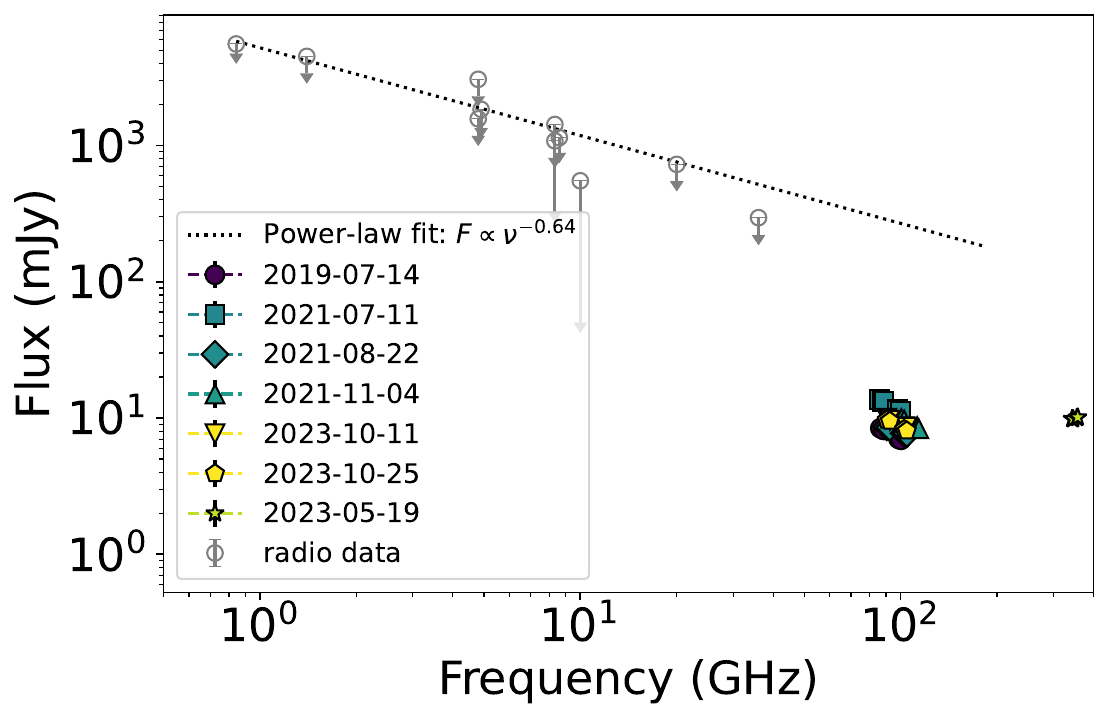}
\caption{NGC~4945}
\label{fig:subim2}
\end{subfigure}

\caption{The compilation of archival radio and ALMA data (see details in Tables~\ref{tab:radio_3783},\ref{tab:radio_mcg},\ref{tab:radio_4945}). The radio data below 100 GHz is fitted with a power law.}
\label{fig:arc_radio}
\end{figure*}

\subsection{The size of the emitting region} \label{sec_size}

The absence of mm variability in any of the sources during our observations suggests that the emitting region is typically larger than one light hour. However, as shown above, the mm observations are too sparse to place stringent constraints on the size of the mm-emitting region in each source.

To derive rough constraints on the emitting region size, we follow the approach of \citet{laor08}, assuming that the mm emission originates from a homogeneous synchrotron source with a uniform magnetic field $B$ and a power-law energy distribution of relativistic electrons. Additionally, we assume that the magnetic energy density is in equipartition with the photon energy density. Under these conditions, the magnetic field strength can be estimated using Eq.~21 from \citet{laor08}:
\begin{equation} \label{b_lb08}
    B_{\rm eq} = 0.27 R_{\rm pc}^{-1} L_{\rm bol}^{1/2} \ \ \  {\rm (G)},
\end{equation}
where $R_{\rm pc}$ is the size of the region in pc, and $L_{\rm bol}$ is the bolometric luminosity in units of 10$^{46}$~erg~s$^{-1}$. We calculated $L_{\rm bol}$ from X-ray luminosity at 14--150~keV (see Table~\ref{tab:sample}) considering a uniform 14--150~keV bolometric correction of 8.48 \citep[following][]{ricci23}. Assuming $B = B_{\rm eq}$, Eq.~19 from \citet{laor08} holds:
\begin{equation} \label{r_lb08}
 R_{\rm pc} = 1.2\times10^{-4} L_{\rm mm}^{0.4} L_{\rm bol}^{0.1} \nu_{\rm GHz}^{-1.4}  \ \ \  {\rm (pc)},
\end{equation}
where $L_{\rm mm}$ is mm luminosity in units of 10$^{30}$~erg~s$^{-1}$, $\nu_{\rm GHz}$ is the observed frequency in GHz.

According to \citet{chen25}, a homogeneous synchrotron source of radius $R$ produces a spectrum composed of self-absorbed emission below the turnover frequency and optically-thin emission above it. This results in a flat spectral index, $0 > \alpha > -0.5$, between two turnover frequencies corresponding to the inner and outer radius of the source: $\nu_0 (R_{\rm max}) < \nu < \nu_0 (R_{\rm min})$. In our sample, the mm emission at 100~GHz exhibits $\alpha < -0.5$ in NGC 3783 and NGC 4945, leading to $R_{\rm min} = 178$ and $586 \, R_{\rm g}$, respectively, based on Eq.~(\ref{r_lb08}). In MCG~5--23--16, $\alpha \approx -0.4$, indicating that the emission may be transitioning between optically-thin and thick regimes. For this source, $R_{\rm min} = 86 R_{\rm g}$ at 100~GHz, though if $\nu_0 (R_{\rm min}) > 100$ GHz, the inner region could be even smaller. According to Eq.~(\ref{b_lb08}), the corresponding magnetic field strengths at $R_{\rm max}$ are $\sim$116~G, $\sim$147~G, and $\sim$111~G for NGC~3783, NGC~4945, and MCG~5--23--16, respectively.

Estimating $R_{\rm max}$ is more challenging, as it requires observations at lower frequencies to identify where the inverted spectrum transitions to a flat one. To address this, we compiled high-resolution radio data from the literature (Fig.~\ref{fig:arc_radio}, see details in Appendix~\ref{radio}). While we aimed to include only data with angular resolution better than 1$''$, for frequencies at or below 1.4~GHz, only data with a resolution of approximately 1$'$ were available. 
In all cases, the low-frequency radio data ($\lesssim$40~GHz) are well described by a power-law fit. However, this fit cannot be extrapolated to the millimeter regime, as the spectral slope at 100~GHz differs significantly. A clear mm excess is observed in NGC~3783 and MCG~5--23--16, strongly deviating from the low-frequency power-law behavior. In contrast, for NGC~4945, the measured radio fluxes are significantly contaminated by surrounding structures. Therefore, we treat all available radio data as upper limits on the intrinsic radio emission of AGN. As a result, the observed radio slope in NGC~4945 likely traces extended emission rather than the compact core.

The mismatch between the radio and mm regimes suggests that the spectral break between optically thick and thin regimes occurs between $\sim$10 and 100~GHz. Based on the radio luminosity detected at the highest frequency that follows the radio power-law, we estimated upper limits for $R_{\rm max}$. Using the flux at 8.6~GHz for NGC~3783, we obtained $R_{\rm max} \approx 3500 R_{\rm g}$, which is comparable to the BLR size \citep{bentz_3783}. Assuming equipartition, the corresponding magnetic field at this $R_{\rm max}$ is $\sim$5~G. For MCG~5--23--16, observations exist at 36 GHz, but the emission already shows an excess compared to lower frequencies. Therefore, we used a flux at 8.46~GHz, yielding $R_{\rm max} \approx 1000 R_{\rm g}$, with a magnetic field strength of $\sim$10~G. In NGC~4945, the situation is more complex due to the lack of data to separate the AGN emission from the surrounding star-forming region. This galaxy exhibits very extended and bright radio emission compared to the others (see Fig.~\ref{radio}). However, this emission is associated with star formation, and there is no clear radio detection of the AGN in NGC~4945.Using a 1~mJy upper limit at 2.3~GHz from \citet{lenc09}, we estimate $R_{\rm max} \approx 10^4R_{\rm g}$, with a corresponding magnetic field strength of $\sim$8~G.

\subsection{Comparison with radio-loud sources} \label{sec_rl}

To compare the polarimetric properties of the RQ AGN in our sample, we examined ALMA polarization data for RL AGN. Blazars and compact radio sources, such as Flat Spectrum Radio Quasars (FSRQs) and Compact Steep Spectrum (CSS) sources, are known to exhibit significant mm polarization, often exceeding 10\%. For example, \citet{3c286} demonstrated that 3C~286, a CSS source with a compact jet, exhibits $\sim$17\% polarization at 230~GHz. Similarly, blazars frequently show polarization levels from a few to more than 10\% (see AMAPOLA database).

To investigate RL AGN with extended radio jets and without strong Doppler beaming, we selected non-blazar RL sources having high-resolution archival ALMA data (<0\farcs5). This dataset included three RL sources: NGC~1052, Cen~A, and 3C~120. Additionally, we considered 3C~273, although it is an FSRQ and therefore not an ideal comparison case. However, \citet{hovatta19} reported that at $\sim$230~GHz, 3C~273 exhibited an unresolved core with a linear polarization of only 1.8\%, attributed to synchrotron emission from the jet base, suggesting that Faraday depolarization may play a significant role.

NGC~1052 was observed multiple times in different bands, consistently revealing a negative spectral slope. The total intensity analysis of Band 4 and 6 data from November 2018 (2018.1.00457.S, PI: S. Kameno) showed a spectral index of $-$0.9 $\pm$ 0.1\footnote{Throughout this section, in-band spectral indices are derived from ALMA archival images in individual spectral windows, assuming flux uncertainties as specified in the ALMA Proposer’s Guide.}. Observations from May--June 2023 (2022.1.00506.S, PI: T. Davis) across Bands 3--6 indicated an even steeper spectrum with a spectral index of $-$1.3 $\pm$ 0.1. Despite this, no polarized signal was detected above the 0.1\% calibration uncertainty in the November 2018 Band 4 and 6 data, confirming the absence of detectable mm polarization.

3C~120 was observed in Bands~4 and 6 (2017.1.01425.S, PI: T. Savolainen) with a 250-day separation. Both observations showed a total intensity spectral slope of about $-$0.5, though the flux varied significantly (by a factor of 1.8 when extrapolated between bands). No polarized signal was detected above the 0.1\% calibration uncertainty.

ALMA observed Cen~A polarization four times (2013.1.01282.S, 2015.1.00421.S, PI: H. Nagai), but only in Band~7, where no polarization above 0.1\% was detected. The total intensity spectra in Bands~3 and 4 (2022.1.00506.S, PI: T. Davis) followed a steep power-law index ($\alpha \approx -$0.3). However, Bands 5 and 7 did not fit this power-law, and the Band 7 flux showed a significant excess, suggesting a possible contribution from another component, such as thermal dust emission.

Overall, despite their optically thin synchrotron emission, the archival ALMA polarimetric data show that RL AGN showed no polarized signal except PD $\sim 1.8$\% in 3C~273 where the polarization level is comparable to the upper limits we obtained for RQ AGN. Faraday depolarization could be a key factor in suppressing polarization, as demonstrated for the jet base of the FSRQ 3C~273 \citep{hovatta19}. Magnetic field geometry may also play a role in reducing the observed polarization. A striking example is M87, where polarization at mm wavelengths reaches 15\% in resolved EHT maps \citep{m87_eht_pol}. However, ALMA observations with 0\farcs3 resolution show much lower polarization: PD $\approx$ 0.5\% in Band 6 (2017.1.00608.S, PI: I. Marti-Vidal), PD $\approx$ 1.3\% in Band 3 (2013.1.01022.S, PI: K. Asada) and PD $\approx$ 2.8\% in Band 7 \citep{goddi25}. These values are comparable to the upper limits obtained for our RQ sample. Notably, our three RQ AGN belong to different types and thus have different inclinations, which, in principle, should mitigate the impact of magnetic field configuration, whether predominantly toroidal or poloidal, on reducing the observed polarization. Nevertheless, no polarization was detected in the RQ sample.

In summary, the few RL AGN with detected mm polarization in archival ALMA data exhibit polarization degrees comparable to the upper limits derived for our RQ sample. The number of the studied sources remains too small to establish a clear trend, however, these findings may suggest that mm PD at the 0.5--1\% level are not clearly indicative of either jet- or corona-dominated emission, at least in the currently available data. Additional insight comes from recent FIR polarimetric studies \citep{lr23}, which reveal a pronounced dichotomy: while radio-quiet AGN appear essentially unpolarized (PD < 1\%), radio-loud AGN show significantly higher polarization (5--11\%, including Cen A with PD $\approx$ 4\% at 100~$\mu$m), attributed to magnetically aligned dust grains at scales of 5--130~pc. This supports the idea that a more strongly magnetized environment with a coherent field structure may be key to jet launching, and further mm polarimetric observations are needed to enable a systematic comparison between RQ and RL AGN.

%-----------------------------------------------------------------

\section{Conclusions}

To reveal the origin and to investigate the properties of the compact mm emission ubiquitously found in RQ AGN, we observed a sample of three mm-bright AGN -- NCG~3783, MCG~5--23--16, and NGC~4945 -- with ALMA Band 3 ($\sim$100~GHz) in full polarization mode. The new observations were deep enough to show in all three galaxies both the compact mm regions located at the position of the AGN and unresolved at $\sim$0\farcs1 angular resolution and faint extended structures surrounding the central mm core (see upper panels in Fig.~\ref{fig:3783}, \ref{fig:mcg}, \ref{fig:4945}). The mm emission from all three sources showed a negative in-band spectral index, which, at least for NGC~3783 and NGC~4945, where $\alpha \approx -1$, is a sign of the optically thin synchrotron emission; MCG~5--23--16 data showed $\alpha \approx -0.4$, which can point at the transition state to the optically thick synchrotron regime.

\begin{itemize}
    \item Despite exhibiting a negative spectral index in the mm band, indicative of optically thin synchrotron emission, no polarization was detected in RQ AGN down to a level of PD = 0.5\%. The complete absence of detected polarization suggests strong depolarization. While depolarization due to a disordered magnetic field alone would require an excessive number of uncorrelated emitting regions, resulting in a smooth flux variability inconsistent with observations, the Faraday effect emerges as a more plausible mechanism. By comparing the expected impact of Faraday rotation with the obtained PD$_{\rm lim}$, we find that Faraday rotation in the X-ray corona is the most likely cause of the total depolarization of the observed mm emission (Sec.~\ref{sec_depol} and \ref{sec_what}). Notably, archival high-resolution ALMA data of central emission in non-blazar RL AGN also show either no mm polarization or one comparable with the upper limits for our RQ sample (Sec.~\ref{sec_rl}).
    \item Deep mm observations revealed faint structures at the $\sim$0.1$-$0.2~mJy/beam level (2--3$\sigma$, <10\% of the AGN emission) surrounding NGC~3783 and MCG~5–23–16, extending over at least a few tens of parsecs, which had not been previously detected (Sec.~\ref{sec_ext}). In NGC~3783, the extended mm emission is co-spatial with the narrow-line region observed earlier by MUSE \citep{denBrok20} and GRAVITY \citep{gravity_3783}.
    \item While no polarization was detected from the AGN itself, NGC~3783 exhibited an unresolved polarized source with a PD $\approx 17$\%, located $\sim$20~pc from the AGN mm core (Sec.~\ref{sec_ext}). Since the separation between the maxima of the total and polarized intensity exceeds the beam size, the polarized source is definitely not the AGN itself. Instead, it is co-spatial with the extended mm structure and the narrow-line outflow. This polarized emission may be linked to the processes in the AGN outflow, such as a propagating shock. The steep in-band spectral index, $\alpha = -3.6 \pm 2.4$, suggests an event with rapid electron cooling; however, the large uncertainty highlights the need for additional observations.
    \item The studied sources show no mm flux variability during the observational sessions, suggesting that the mm source size exceeds$\sim$1~light~hour. A compilation of archival ALMA Band 3 data for the RQ AGN in our sample revealed that the mm flux varies in all sources by a typical factor of 2 (Sec.~\ref{sec_var}). However, due to the lack of regular mm monitoring and simultaneous observations in other bands, we cannot determine the mechanism driving this variability. Assuming energy equipartition, we estimated the minimum size of the emitting synchrotron sources to be $\sim$90--600~$R_{\rm g}$, while archival radio data provided upper limits on the emitting region size of $10^3 - 10^4$~$R_{\rm g}$ (Sec.~\ref{sec_size}).
\end{itemize}

From these polarimetric ALMA observations, the key insight is not the detection of mm polarization itself, but rather its absence with a very low upper limit of 0.5\%, providing a crucial clue to the origin of the compact mm emission. Among various synchrotron-emitting AGN structures, only the X-ray corona, because of its electron density, exhibits sufficient Faraday depolarization to fully suppress the observed polarization. However, to strengthen this argument, highly sensitive mm polarimetric observations with higher spectral resolution, on the order of at least a few tens of MHz, are necessary to avoid bandwidth depolarization caused by the large rotation measures discussed in Sec.~\ref{sec_depol}. Additionally, extending observations to higher frequencies, up to $\sim$500~GHz, would help probe deeper into the emitting region.

\begin{acknowledgements}
This paper makes use of the following ALMA data: ADS/JAO.ALMA\#2023.1.01517.S. ALMA is a partnership of ESO (representing its member states), NSF (USA) and NINS (Japan), together with NRC (Canada), NSTC and ASIAA (Taiwan), and KASI (Republic of Korea), in cooperation with the Republic of Chile. The Joint ALMA Observatory is operated by ESO, AUI/NRAO and NAOJ.
 We sincerely thank Ari Laor for the valuable comments and feedback.
 ES acknowledges ANID BASAL project FB210003 and Gemini ANID ASTRO21-0003. CR acknowledges support from Fondecyt Regular grant 1230345, ANID BASAL project FB210003 and the China-Chile joint research fund.
\end{acknowledgements}

\bibliographystyle{aa} % style aa.bst
\bibliography{biblio} % your references Yourfile.bib

\begin{appendix}

\section{ALMA data}

The tables provided contain the data used in Fig.~\ref{fig:arc_alma}.

\begin{table}[!ht]
\centering
\caption{Archival ALMA Band 3 data for NGC~3783. $\alpha$ is the in-band spectral index, and $\theta$ is the angular resolution.}
\begin{tabular}{lcccc}
\hline
Date       & Flux  & rms & $\alpha$   & $\theta$      \\
             & \footnotesize{(mJy)} &  \footnotesize{(mJy/beam)} & &             \\ \hline
2019--08--13 & 2.19  & 0.04 & $-$1.34 $\pm$ 0.39 & 0\farcs3 \\
2019--08--18 & 2.55  & 0.04 & $-$0.26 $\pm$ 0.16 & 0\farcs3\\
2021--08--21 & 2.16  & 0.03 & $-$0.91 $\pm$ 0.35 & 0\farcs1\\
2021--11--07 & 2.51  & 0.02 & $-$0.49 $\pm$ 0.19 & 0\farcs2\\
2023--05--30 & 3.91  & 0.02 & $-$0.56 $\pm$ 0.06 & 0\farcs3\\
2023--10--13 & 2.03  & 0.01 & $-$0.92 $\pm$ 0.11 & 0\farcs1 \\ \hline
\end{tabular}
\label{arc_alma_3783}
\end{table}

\begin{table}[!ht]
\centering
\caption{Archival ALMA Band 3 data for MCG~5--23--16. $\alpha$ is the in-band spectral index, and $\theta$ is the angular resolution.}
\begin{tabular}{lcccc}
\hline
Date         & Fux  & rms & $\alpha$   & $\theta$      \\
             & \footnotesize{(mJy)} &  \footnotesize{(mJy/beam)} & &             \\ \hline
2021--08--14  & 4.12  & 0.03      & 0.29 $\pm$ 0.04      & 0\farcs1        \\
2021--11--04  & 3.95   & 0.02     & 0.32 $\pm$ 0.13     & 0\farcs2         \\
2023--10--08  & 2.53   & 0.01     & $-$0.42 $\pm$ 0.04   & 0\farcs1          \\ \hline
\end{tabular}
\label{arc_alma_MCG}
\end{table}

\begin{table}[!ht]
\centering
\caption{Archival ALMA Band 3 data for NGC~4945. $\alpha$ is the in-band spectral index, and $\theta$ is the angular resolution.}
\begin{tabular}{lcccc}
\hline
Date       & Flux  & rms & $\alpha$   & $\theta$      \\
             & \footnotesize{(mJy)} &  \footnotesize{(mJy/beam)} & &             \\ \hline
2019--07--14 & 7.60  & 0.01 & $-$1.35 $\pm$ 0.03  & 0\farcs1 \\
2021--07--11 & 13.17 & 0.03 & $-$1.34 $\pm$ 0.03  & 0\farcs3 \\
2021--08--22 & 8.14  & 0.03 & $-$0.84 $\pm$ 0.06  & 0\farcs1 \\
2021--11--04 & 9.41 & 0.22 & $-$1.34 $\pm$ 0.03  & 0\farcs2 \\
2023--10--11 & 9.12 & 0.01  & $-$1.06 $\pm$ 0.08  & 0\farcs1 \\
2023--10--25 & 8.88 &  0.01 & $-$1.29 $\pm$ 0.04  & 0\farcs1 \\
\hline
\end{tabular}
\label{arc_alma_4945}
\end{table}

\section{Radio data} \label{radio}

The tables provided contain the data used in Fig.~\ref{fig:arc_radio}.

\begin{table*}[]
\centering
\caption{Archival radio data for NGC~3783.}
\begin{tabular}{lcccc}
\hline
Frequency & Flux & Flux error & Resolution & Reference  \\
(GHz)      & (mJy)      & (mJy)      & (arcsec)       &            \\ \hline
0.843      & 58.4       & 2.1        & 60             & \citealt{sumss}     \\
1.49       & 43.6       & 2.0        & 45             & \citealt{nvss}        \\
1.49       & 23.0       & 2.0        & 0.6            & \citealt{Unger87}    \\
4.86       & 13.0       & 1.3        & 0.6            & \citealt{Unger87}    \\
4.9        & 13.0       & 1.0        & 0.61           & \citealt{Ulvestad84} \\
8.6        & 10.03      & 0.5        & 1.59 $\times$ 0.74    & \citealt{Morganti99} \\
8.6        & 8.0        & 0.4        & 0.25           & \citealt{Schmitt01}  \\ \hline
\end{tabular}
\label{tab:radio_3783}
\end{table*}

\begin{table*}[]
\centering
\caption{Archival radio data for MCG~5--23--16.}
\begin{tabular}{lcccc}
\hline
Frequency & Flux & Flux error & Resolution & Reference  \\
(GHz)      & (mJy)      & (mJy)      & (arcsec)       &            \\ \hline
1.4        & 11.0       & 1.5        & 0.61           & \citealt{Ulvestad84} \\
4.8        & 6.0        & 1.0        & 0.61           & \citealt{Ulvestad84} \\
8.46       & 2.6        & 0.13       & 0.43 $\times$ 0.23      & \citealt{Orienti10}  \\
22         & 3.176      & 0.159      & 1              & \citealt{magno24} \\ \hline
\end{tabular}
\label{tab:radio_mcg}
\end{table*}

\begin{table*}[]
\centering
\caption{Archival radio data for NGC~4945. Note that, regardless of resolution, the flux was integrated over the entire central emitting region of the galaxy.}
\begin{tabular}{lcccc}
\hline
Frequency & Flux   & Flux error & Resolution    & Reference                  \\
(GHz)     & (mJy)  & (mJy)      & (arcsec)      & \multicolumn{1}{l}{} \\ \hline
0.843     & 5549   & 166.5      & 60            & \citealt{sumss}                \\
1.4       & 4500   & 225        & 45            & \citealt{nvss}                 \\
4.8       & 1566   & 78         & <0.5              & \citealt{burlon13}             \\
4.8       & 3055   & 153        & $\sim$1      & \citealt{CRATES}               \\
4.9       & 1840   & 100        & 1.2 $\times$ 1.1     & \citealt{Elmouttie97}          \\
%5         & 1897   & 31         & 6.1 $\times$ 4.6     & \citealt{McConnell12}          \\
8.33      & 1424   & 220        & <1            & \citealt{Roy10}                \\
8.33      & 1080   & 54         & $\sim$1      & \citealt{CRATES}               \\
%8.6       & 786    & 39         & <0.5             & \citealt{burlon13}             \\
8.6       & 1141   & 52         & 5.6 $\times$ 4.2     & \citealt{McConnell12}         \\
10        & 550    & 100        & 0.7 $\times$ 0.6     & \citealt{Elmouttie97}         \\
20        & 726    & 36         & <0.5           & \citealt{burlon13}             \\
36        & 295    & 15         & 5 $\times$ 2         & \citealt{mccarthy18}           \\ \hline
\end{tabular}
\label{tab:radio_4945}
\end{table*}

\end{appendix}

\end{document}